\documentclass[aps,english,prb,showpacs,twocolumn,superscriptaddress]{revtex4}
\usepackage{graphicx}
\usepackage{amssymb}
\usepackage{babel}

\begin{document}

\title{Domain formation on oxidized graphene}

\author{M. Topsakal}
\affiliation{UNAM-National Nanotechnology Research Center, Bilkent University, 06800 Ankara, Turkey}
\affiliation{Institute of Materials Science and Nanotechnology, Bilkent University, Ankara 06800, Turkey}
\author{S. Ciraci}\email{ciraci@fen.bilkent.edu.tr}
\affiliation{UNAM-National Nanotechnology Research Center, Bilkent University, 06800 Ankara, Turkey}
\affiliation{Institute of Materials Science and Nanotechnology, Bilkent University, Ankara 06800, Turkey}
\affiliation{Department of Physics, Bilkent University, Ankara 06800, Turkey}

\date{\today}

\begin{abstract}

Using first-principles calculations within density functional theory we demonstrate that the adsorption of single oxygen atom results in significant electron transfer from graphene to oxygen. This strongly disturbs the charge landscape of the C-C bonds at the proximity. Additional oxygen atoms adsorbing to graphene prefer always the C-C bonds having highest charge density and consequently they have tendency to form domain structure. While oxygen adsorption to one side of graphene ends with significant buckling, the adsorption to both sides with similar domain pattern is favored. The binding energy displays an oscillatory variation and the band gap widens with increasing oxygen coverage. While a single oxygen atom migrates over the C-C bonds on graphene surface, a repulsive interaction prevents two oxygen adatoms from forming an oxygen molecule. Our first-principles study together with finite temperature ab-initio molecular dynamics calculations concludes that oxygen adatoms on graphene cannot desorb easily without influence of external agents.
\end{abstract}

\maketitle

\section{Introduction}

Graphene, strictly two dimensional allotrope of carbon atom with its unique
mechanical\cite{mechanical}, structural\cite{structural}, electronic\cite{electronic1,electronic2}
and thermal properties,\cite{thermal} has been considered as a promising candidate
for next generation electronic devices and numerous nanoscale applications. Ingenious
methods have been proposed for its production.\cite{prod1,prod2,dai,iijima,tongay,taner}
Intensive studies have been also carried out for controlling and modifying various
properties of bare graphene. The adsorption of foreign atoms or molecules
on bare graphene surface has been considered as an efficient method to attain this
objective.

Graphene oxide (GOX) is an  example\cite{gox1,gox2} to show how the properties of graphene
can be changed dramatically upon the adsorption of oxygen atoms. GOX is obtained through
oxidative exfoliation of graphite, which can be visualized as an individual sheet of graphene
decorated with epoxy (C-O-C) and hydroxyl (C-OH) groups on both sides and edges. Incidentally,
GOX has been also an attractive material for large scale graphene production\cite{aksay} due
to low-cost, simple and high yield reduction methods. Unfortunately, despite the oxidized
graphite is a known material since last 150 years\cite{1800s} and great deal of experimental
and theoretical research carried out recently,\cite{li,boukhalov_2008,yan_2009,yan-prb,akbar,wang_2010,xiang,sun,mattson,kim_nature}
a thorough understanding regarding the interaction of oxygen with graphene and relevant reactions
are not available yet due their stochastic nature.

\begin{figure}
\includegraphics[width=8cm]{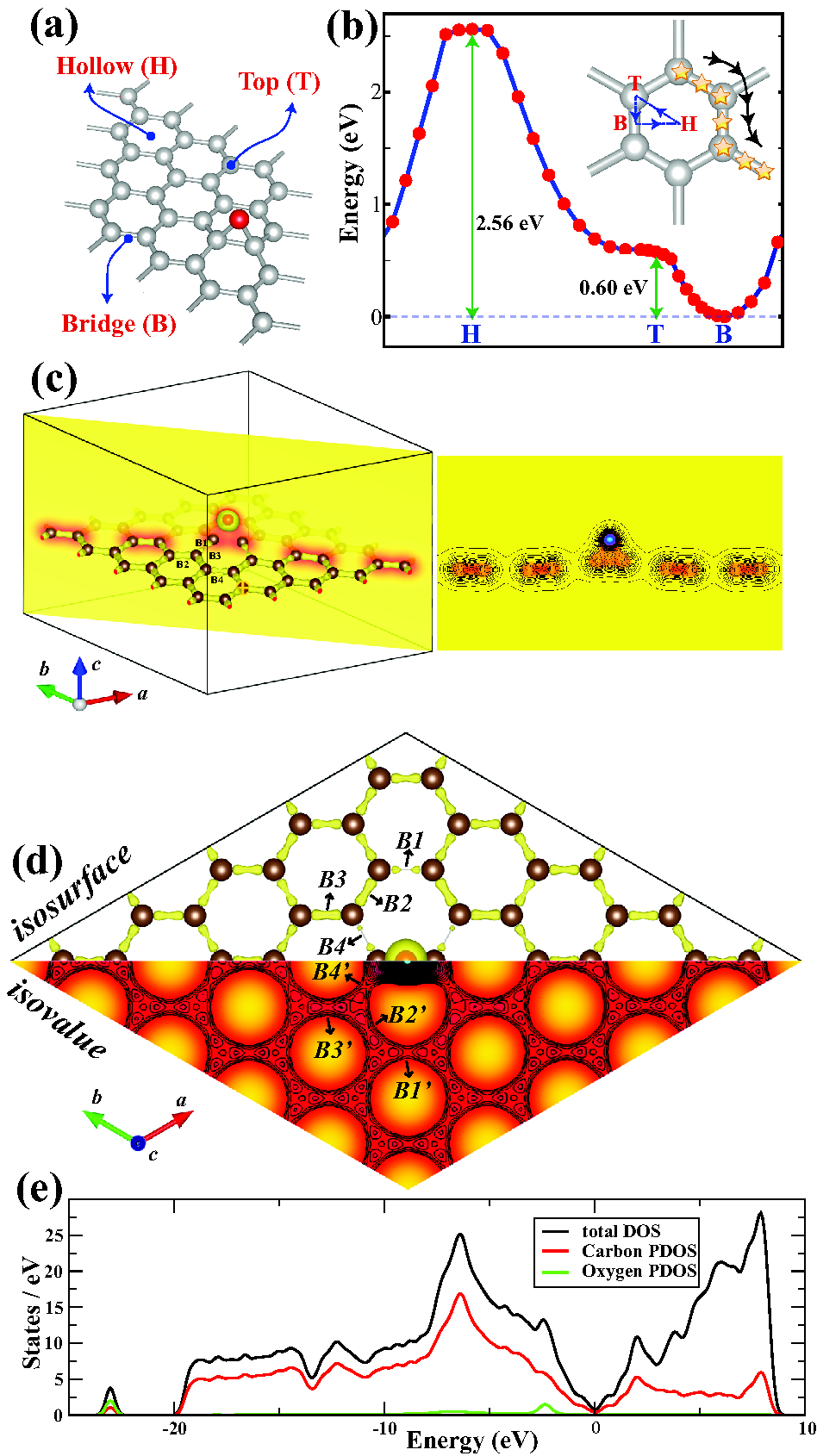}
\caption{Various critical sites of adsorption on the 2D honeycomb structure of graphene and an
oxygen atom adsorbed on the bridge site, which is found to be as the energetically
most favorable site. Carbon and oxygen atoms are shown by gray and red balls, respectively.
(b) Variation of energy of oxygen adatom adsorbed to graphene along $H \rightarrow T \rightarrow B$ directions of the hexagon. The diffusion path of a single oxygen adatom is shown by stars. (c) Charge density isosurfaces, isovalues and contour plots of oxygen adsorbed graphene in a plane passing through C-O-C atoms. (d) Same as (c) on the lateral plane of honeycomb structure. (e) Total and partial density of states projected
to carbon and oxygen atoms. Calculations are carried out for supercell presented in (c) where O-O interaction is significantly small.}
\label{Figure1}
\end{figure}

To understand the experimental data, various structural configurations of GOX
have been proposed based on first principles calculations. Performing the analysis of various coverage models, Boukhalov \textit{et al.}\cite{boukhalov_2008} revealed that 100\% coverage of GOX is energetically less favorable than 75\% coverage. Also, while a coverage less than 25\% of GOX contains only hydroxyl groups, the mixed GOX consisting of both oxygen and hydroxyl is favored for higher coverage. In a later study, Yan \textit{et al}.\cite{yan_2009} suggested that it is energetically favorable for the epoxy and hydroxy groups to aggregate together to form specific types of strips with $sp^2$ carbon regions in between. In contrast, Wang \textit{et al.}\cite{wang_2010} argued that thermodynamically stable structures are fully covered without any $sp^2$ carbon. The domains of graphene monoxide with $N_{O}/N_{C}=$1 (\textit{i.e.} the ratio of number of oxygen $N_O$ to the number of carbon atoms $N_C$) is attained by the oxidation of both sides.\cite{mattson} Very recent study\cite{kim_nature} combining experimental results and first principles calculations shows that multilayer GOX is metastable at room temperature undergoing modifications and reduction with a relaxation time of approximately 35 days. At the quasi-equilibrium, the nearly stable oxygen coverage  was reported as $\Theta$=0.38 and presence of C-H species is found to favor the reduction of epoxides and to a lesser extent hydroxyl groups with the formation and release of water molecules.\cite{kim_nature}

From our point of view, there exists still controversies between theory and experiment.
For example, yet the distribution of hydroxy and epoxy groups on GOX surface together
with the trends related with their clustering or uniform coverage are unknown.
At least, a rigorous explanation for the reason of the differences in the interpretations
of experimental data is required. In particular, it is not clear why the desorption of
oxygen adatoms through O$_2$ formation does not occur so easily despite the negative
formation energy of oxygen adsorption. Unlike GOX, the hydrogenated graphene, i.e.
graphane (CH)\cite{CH1,CH2} and fluorinated graphene, i.e. fluorographene
(CF)\cite{CF1,CF2} are experimentally realized and their crystal structure are well understood.

In this study we present an extensive analysis of the oxygen adsorption and oxygen coverage
by using first principles calculations based on Density Functional Theory (DFT). In order to understand
the reversible oxidation-deoxidation processes\cite{science,gox2} we consider only oxygen adatoms
on graphene surfaces, in spite of the fact that hydroxyl groups are readily coadsorbed. Earlier
studies have followed approaches, which consider the optimized geometries
corresponding to the minimum of total energy. Here, we show that the mechanism of oxygen coverage is
governed mainly by the charge density profile of graphene, which is modified by each adsorbed oxygen
in the course of oxidation. At the end, oxidized regions of graphene tend to form domains instead of a
uniform coverage. In view of these results we also discuss unzipping process of graphene.\cite{aksay,sun}
The oxygen adsorption
on both sides of graphene was shown to be energetically more favorable than the adsorption to only one side,
whereby serious distortions of the graphene lattice occurred. The repulsive interaction between two oxygen
adatoms at the close proximity is repulsive and hinders oxygen desorption through O$_2$ formation.
We finally showed that the distribution of oxygen atoms on graphene affects the electronic properties.
Even if the massless Dirac-fermion behavior of graphene can be recovered for patterns conserving
specific symmetries, the band gap normally increases with increasing non-uniform oxygen coverage
and attains the value as high as 3 eV. These results are critical for the device applications based
on reversible oxidation-deoxidation of graphene surfaces.\cite{science,gox2}

\section{Method}

Calculations are carried out within spin-polarized and spin-unpolarized
density-functional theory (DFT) using  projector augmented wave
(PAW) potentials.\cite{paw} The numerical calculations have been
performed by using VASP package.\cite{vasp1,vasp2} The exchange correlation
potential is approximated by generalized gradient approximation functional
of Perdew, Burke, and Ernzerhof (PBE).\cite{pbe} Calculations are carried out
using periodically repeating supercell geometry, where the spacings between
graphene layers are taken 15 \AA. However, systems involving very large graphene
sheets are treated with 10 \AA~spacing, which is still large and hinders interlayer
coupling. A plane-wave basis set with kinetic energy cutoff of 500 eV is used. All
atomic positions and lattice constants are optimized by using 
the conjugate gradient method, where the total energy and atomic
forces are minimized. The convergence for energy is chosen as 10$^{-5}$
eV between two steps. Oxygen-adatom and graphene system breaks
inversion symmetry and a net electric-dipole moment is generated
perpendicular to the graphene surface. Dipole corrections\cite{dipole}
are applied in order to remove spurious dipole interactions between
periodic images. The $\Gamma$-point i.e. \textbf{k}=0 is used for rectangular
supercells containing 128 carbon atoms and oxygen adatoms, while 18x18x1
\textbf{k}-point sampling is used for primitive unit-cell. The Gaussian
smearing with a width of 0.1 eV is used in the occupation of electronic energy bands.

\section{ Interaction of Oxygen Atom with Graphene}

A thorough analysis of the interaction of single O atom with graphene is essential
to understand the oxidation process. Here the adsorption of single (isolated) oxygen
on graphene is represented using large supercells, where O-O interaction is minimized.
Owing to its hexagonal crystal structure, there are three major sites for foreign 
atom adsorption on graphene as shown in Fig. \ref{Figure1} (a). The hollow (H) site is
above the center of hexagonal rings formed by carbon atoms. The top (T) site lies on top
of the carbon atoms and the bridge (B) site is above the middle of each bonds
connecting two adjacent carbon atoms. The bridge site is found to be most favorable
adsorption site for an oxygen atom. Earlier LDA calculations predicted also B-site
as energetically favorable site.\cite{a-nakada} The variation of the total energy
along H $\rightarrow$ T $\rightarrow$ B sites is presented in Fig. \ref{Figure1} (b).
The energy barrier is 0.6 eV for an O atom diffusing from bridge to top site
and the energy difference between bridge and hollow site is as high as 2.56 eV. Therefore,
the migration paths of oxygen adatom with minimum energy barrier follow the honeycomb
structure over the C-C bonds by going from B- to T-sites as illustrated by inset in
Fig. \ref{Figure1} (b). On the other hand, the energy barrier against the penetration of an
oxygen adatom from one side of graphene to the other side is as high as 6 eV.\cite{coating}
This high energy barrier suggests that graphene can be used an ideal coating preventing
surfaces from oxidation. We note that the hollow site of graphene is more favorable for other
atoms\cite{can_li,haldun_tm} such as Li or Ti, while H and F atoms prefers the top site
for adsorption.\cite{CH2,CF2}

The binding energy of oxygen on graphene is defined as

\begin{equation}\label{equ:binding}
E_b=E_{T}[Gr] + E_{T}[O] - E_{T}[Gr+O]
\end{equation}

where $E_{T}[Gr+O],E_{T}[Gr],E_{T}[O]$ denote the optimized total energies of graphene with
adsorbed oxygen, pristine graphene and free O atom, respectively. Our calculations
show that $E_b = 2.35$ eV for the (2x2) graphene supercell containing 8 carbon and one oxygen atom,
but it increases to 2.40 for (3x3) and to 2.43 for (4x4) supercells. For supercells larger than (4x4),
which correspond to smaller oxygen coverage and hinders O-O coupling, the binding energy does not
change and mimics the binding energy of single, isolated oxygen attached to graphene surface. The calculated
binding energy for full coverage $\Theta$=0.5 (namely the ratio, $N_C$, $N_{O}/N_{C}$=0.5) is 2.80 (3.34) eV per oxygen atom for one-sided (two-sided) adsorption. The binding energy of single oxygen adatom increasing from 2.43 eV to 2.80
eV at full coverage indicates a significant O-O coupling. We note that the formation energy $E_{f}=E_{b}-E_{b,O_{2}}/2$, where $E_{b,O_{2}}$ is the binding energy of O$_2$ molecule) is 
negative for one-sided coverage indicates instability. However, this situation does not impose 
desorption of O through O$_2$ formation for reasons discussed in Sec. V.

According to the Pauling scale, oxygen has an electronegativity of 3.44, which is the second
highest in periodic table after fluorine (3.98) and hence the oxidation of graphene is
expected to result in significant charge transfer between oxygen and carbon atoms. Our
calculations using Bader analysis\cite{bader} estimates a charge transfer of 0.79
electrons from carbon atoms of graphene to oxygen. This charge is mainly transferred
from the nearest two carbon atoms forming the bond above which the oxygen adsorbed
at bridge site, while some nearby oxygen atoms also contribute to the charge transfer.
Figure \ref{Figure1} (c) shows isosurface and isovalue (contour) plots of total
charge for a plane passing through carbon atoms and oxygen. The direction of
electron density increasing from carbon atoms towards oxygen atom is a clear
indication of charge transfer. In addition, two carbon atoms below oxygen
are slightly raised from the plane of other carbon atoms and the charge distribution
of the bond between them is disturbed.

In Fig. \ref{Figure1} (d) bird's-eye view of isosurface and isovalue of charge density
profile of a single oxygen adsorbed to each (5x5) supercell of graphene is presented.
The structure is symmetric  and the oxygen atom is at the center. We label some
of the bonds corresponding to nearby bridge sites as B1,B2,B3,B4 and denote their
equivalent sites by primes. While the isosurface plots of all C-C bonds of bare graphene
are identical, the adsorption of single oxygen modifies the charge distribution
at its close proximity. In fact, the isosurfaces plotted for 0.3 (electron/\AA$^{3}$
)\cite{vesta} show that the electron population of specific bonds are higher.
As clearly seen from the figure, B2 and B3 bonds contain more electrons than in B1
and B4 bonds. The reason for the electron depletion in these bonds is related
with the donation of electrons from these bonds to adsorbed oxygen. Interestingly,
the bonds B2, B3 and their four images contain more electron density compared to B1
and other bonds further away from the oxygen atom. For a better illustration of
bond charge alternation, we also presented the isovalue plot of total charge density
in the upper triangle of Fig. \ref{Figure1} (d). Again, the more isolines in the
isovalue map corresponds to more charge at B2' and B3' compared to B1' and B4'. In
this context, we note that the long range interactions and Friedel oscillations
found in 1D carbon chain\cite{longrange} and 2D graphene\cite{bacsi} induced by
adatoms.  Finally we include the density of states (DOS) plot in Fig. \ref{Figure1} (e)
for the system presented in Fig. \ref{Figure1} (a). The overall total DOS represents
a profile similar to bare graphene DOS making a dip near the Fermi level corresponding
to Dirac points and DOS projected to oxygen atom is represented by a peak around -2.5
eV. Later we show that the electronic density and band gap will change with oxygen coverage.

\subsection{ Interaction of Single Oxygen Molecule ($O_2$) with Graphene}
In contrast to oxygen atom, an oxygen molecule has a weak binding with graphene.
We calculated its binding energy to be 115 meV which consists of 57 meV Van der Waals
interaction\cite{grimme} and 58 meV chemical interaction. Its magnetic moment is 1.90 $\mu_B$,
slightly smaller than the magnetic moment of free O$_2$ due to weak chemical interaction.
The $O_2$ molecule lies parallel, approximately 3 \AA{} above the graphene plane, and
do not induce any distortions to graphene honeycomb structure as in the case of single
oxygen adsorption. Accordingly, the binding of O$_2$ to graphene is specified as
physisorption. It is therefore concluded that graphene cannot be oxidized directly by
O$_2$ molecule unless its dissociation into oxygen atoms takes places at the vacancy
site.\cite{coating}

\section{Coverage of graphene surface with oxygen atoms}

\subsection{Coverage of oxygen on one side}

\begin{figure*}
\includegraphics[width=17.5cm]{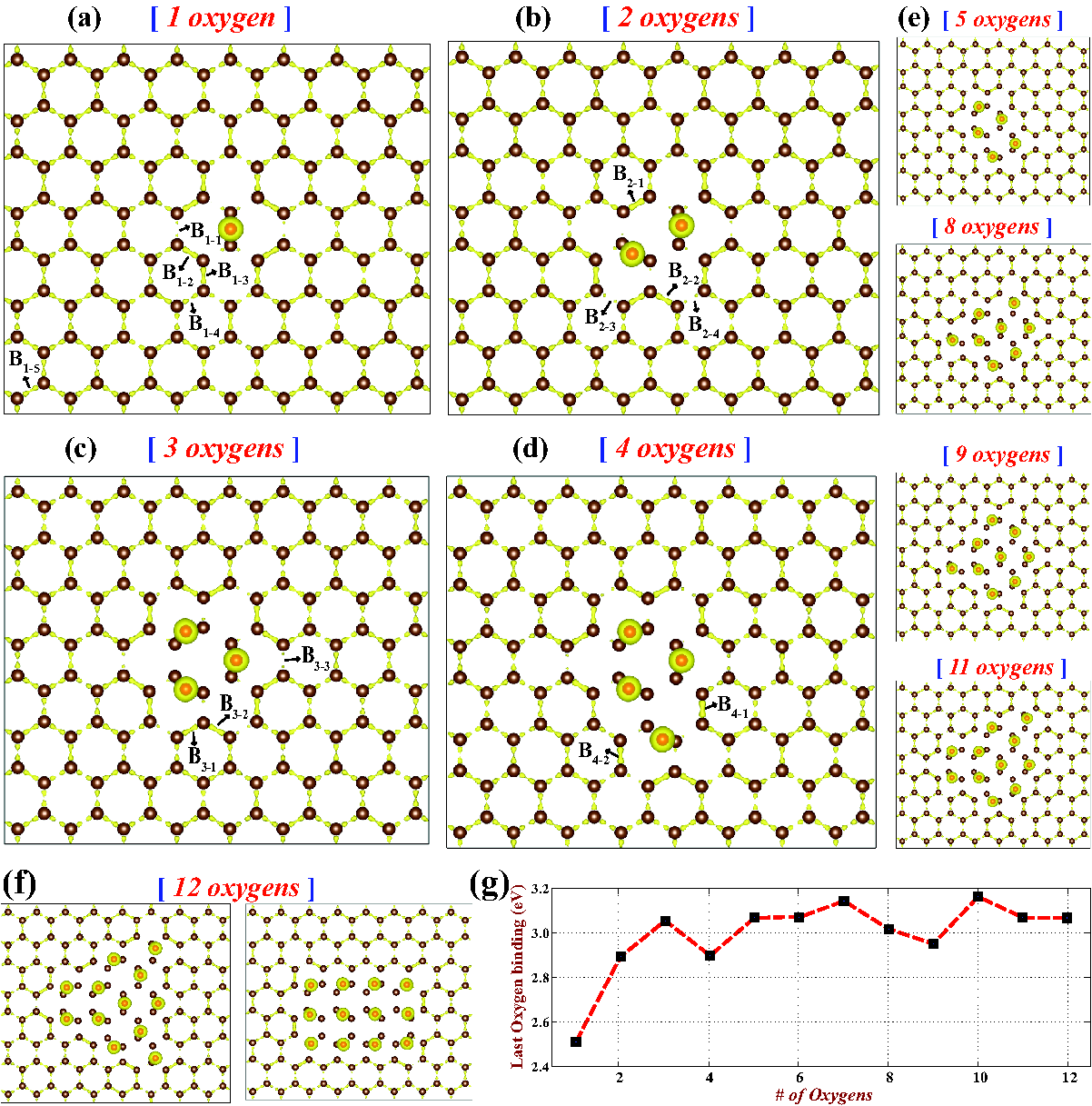}
\caption{(Color Online) (a) Charge density isosurfaces in a
rectangular supercell containing 128 carbon atoms and a single adsorbed O atom shown by a red dot. B$_{i,j}$ identifies a specific C-C bond, where $i$ indicates the total number of oxygen atoms in the supercell and $j$ labels some of the bond around adsorbed oxygen atom(s). (b)-(c) and (d) are same as (a), except that 2,3, and 4 oxygen atoms are adsorbed to the sites, which are most favorable energetically. (e) Energetically favorable configurations up to 11 oxygen atoms adsorbed on graphene deduced from charge. (f) Energetically favorable configuration (left) and less energetic, ordered configuration (right) for 12 O atoms. (g) Variation of the binding energies of the last oxygen adatom up to 12.}

\label{Figure2}
\end{figure*}

Starting from single oxygen adatom, we next consider the adsorption of more
oxygen atoms one at a time on graphene surface leading to higher coverage of oxygen.
We exclude the hydroxyl groups in the present study to simplify the situation and
hence to reveal essential aspects of oxygen adsorption.
In order to reduce the effects of cell size, we construct a larger rectangular
supercell containing 128 carbon atoms as shown in Fig. \ref{Figure2} (a). The
isosurface charge density profile for rectangular supercell is similar to the
charge density profiles in Fig. \ref{Figure1} (d) with $B_{1-2}$ and $B_{1-3}$ bonds having
more charge compared to $B_{1-1}$ and $B_{1-4}$. For the adsorption of second
oxygen, we try all inequivalent bridge sites and calculate their total energies.
It turns out that, the energetically most favorable site for the second oxygen
adsorption is at $B_{1-2}$ site in Fig. \ref{Figure2} (a). In addition, the calculated
binding energy of the second oxygen is around 2.9 eV and this is even higher than
the binding energy of single oxygen on graphene. The binding energy is 154 meV lower
for adsorption on $B_{1-3}$. The binding energy at $B_{1-5}$ site is equal
to the single oxygen binding energy. But interestingly, $B_{1-1}$ and $B_{1-4}$
sites are energetically less favorable sites for second oxygen adsorption compared to other sites.
The calculated $E_{b}$ is 2.33 eV for $B_{1-1}$ site. These calculated energies
indicate a direct correlation between the binding energies and isosurface profiles 
given in Fig. \ref{Figure2} (a). Apparently an oxygen atom prefers the bridge
sites, where the electron density is highest compared to other available sites.

The charge density isosurface profile of graphene supercell containing two
oxygen atoms is presented in Fig. \ref{Figure2} (b) and this profile can
be used to predict the energetically favorable and unfavorable sites for
the adsorption of third oxygen. Again, there are some bridge sites such
as $B_{2-1}$ and $B_{2-2}$ containing more electronic charge compared to
other bonds like $B_{2-3}$ and $B_{2-4}$. The third oxygen is bound to $B_{2-1}$ site
with $E_{b} = 3.06$ eV which is slightly higher for the maximum binding
energy of second oxygen. The binding energies at $B_{2-3}$ and $B_{2-4}$
are approximately 0.7 eV smaller than the binding energy at $B_{2-1}$ site. For
the case of fourth oxygen, $B_{3-2}$ site in Fig. \ref{Figure2} (c) having
more bond charge compared to other sites is energetically most favorable.
It's binding energy,  $E_{b} = 2.90$ eV, is slightly smaller than the binding
energy of previous oxygens. The favorable binding energy for fifth oxygen can be
predicted as $B_{4-1}$ from Fig. \ref{Figure2} (c).

The oxidation process of graphene for more than four oxygen is presented
in Fig. \ref{Figure2} (e) and (f) up to 12 oxygens adatoms. The main trend is that
each oxygen added to system prefers the bridge sites containing higher bond
charge. For the sake of comparison, we included the ordered configuration for 12 adatoms in Fig. \ref{Figure2} (f). However, this configuration (right) is significantly less energetic, by 1.46 eV compared to to the random configuration on the left. We continue to examine the growth of the domain consisting of 12 atoms by adding 
oxygen atoms to the system. The 13$^{th}$ oxygen inserted to the system (not shown in figure) prefers the bridge site on the bond having highest electronic charge, but not the third bridge sites stacking
eventually three oxygen atoms along a line of bridge sites of consecutive 
parallel C-C bonds. There are two such possible sites in Fig. \ref{Figure2}, which are 
identified as the precursors of unzipping (where the usual angle of C-O-C bridge bond increases
by breaking (or weakening) the C-C bond underneath,\cite{aksay,sun} are energetically unfavorable  by $\sim$0.9 eV. The 14th oxygen atom behaved like the previous one: instead of occupying two possible  sites of precursor states, it is adsorbed to a different bridge site which is energetically 614 meV more favorable. 

Clearly, the final structure is a domain of oxygen adatoms on graphene for $\Theta <$ 0.5 
and hence it lacks the signatures of any uniform coverage which is present for the
case of hydrogen and flourine adsorption on graphene. In the case of oxygen, adatoms arrange 
themselves on graphene starting from a single adatoms. Subsequently, additional ones seek 
energetically most favorable sites clustering around the existing ones. This domain structure 
The binding energies of the last adsorbed oxygen atom (or $n^{th}$ adsorbed oxygen) is 
calculated from the expression $E_{b}[n] = (E_{T}[n-1] + E_{T}[O]) - E_{T}[n]$ in terms 
of the minimum total energies of $n-1$ and $n$ oxygen atoms adsorbed on the same supercell of graphene.
For any $n$, the lowest total energy (with negative sign) $E_{T}[n]$ and hence highest binding energy
(with positive sign) $E_{b}[n]$ is determined by comparing the calculated energies of $n^{th}$ 
oxygen adatom when adsorbed to sites remained from the domain of adsorbed $(n-1)$ oxygen atoms.
Once oxygen adatoms nucleate a domain they prefer to grow it by including additional oxygen
atoms, whereby uniform oxidation of graphene surface is precluded.
This conclusion is attained by calculating the total energy of a single domain consisting of 12 
oxygen atoms formed on a large graphene supercell consisting 256 carbon atoms and comparing it 
with the total energy of two separate domains of 6 oxygen atoms nucleated at two different 
locations on the same supercell. The growth of a single domain is found to be favored 
by 330 meV compared to the growth two separate domains. It should be noted that the present 
analysis is done under the condition, where sequential adsorption of oxygen adatoms achieved
in equilibrium. However, oxidation is a stochastic process comprising processes or events taking
place in nonequilibrium. Therefore, growth of multiple domains at finite temperature may occur, but the
uniform growth appears to be a case of least probability. 

In  Fig. \ref{Figure2} (g), the calculated binding energy, $E_{b}[n]$ exhibits an oscillatory variation for $n >2$ and $E_{b}[n=12]$=3.1 eV. These oscillations are physical since energetically favorable site for the $n$th adsorbed
oxygen and the resulting $E_{b}$ are well-determined and unique, but not equivalent to previous sites. The oscillations of $E_{b}$ originate from the changes of the charge distribution and distortions of C-C bonds on the graphene surface occurred as a result of adsorbed oxygen atoms forming a domain.

At this point, we note that the energetic sites of two and three oxygen adatom in Fig. \ref{Figure2} is in agreement with the adsorption sites found by Yan and Chou\cite{yan-prb} as well as by Sun and Fabris.\cite{sun} However, the ground state configuration of freely adsorbed oxygen atoms forming a domain comprising four or more atoms in Fig. \ref{Figure2} are different from that leading to the C-C unzipping, since the latter configuration first require to overcome a significant energy barrier.\cite{sun} As a matter of fact, two oxygen adsorbed to the bridge sites on two parallel C-C bonds of the same hexagon was unfavorable energetically by 1.5 eV.\cite{yan-prb,sun} On the other hand, at advanced stages of Fig. \ref{Figure2} comprising a domain of 8 oxygen atoms we obtained a configuration consisting two B-sites occupied by oxygen atoms on the adjacent parallel C-C bonds, which can be a precursor of unzipping if subsequently adsorbed oxygen atoms occupy additional adjacent B-sites on a line and they overcome an energy barrier to increase C-O-C bond by breaking the C-C bond underneath. However, next adsorbed oxygen as  well as 13$^{th}$ and 14$^{th}$ stopped to develop the precursor state and preferred different sites. We note that at finite temperature and in the presence of other external effects oxygen atoms are prone to develop precursors of unzipping by deviating fro above sequence of adsorption taking place in equilibrium. Later in Sec. VII we discuss this issue further. We also note that patterns of  oxygen coverage for $N_{O}/N_{C}$=0.5 predicted by the genetic algorithm\cite{xiang} cannot be directly comparable with the nonuniform coverage in the present study, which determines the most energetic sites when oxygen atoms are adsorbed on graphene sequentially one at a time.

\subsection{High temperature behavior}

In order to test the stability of oxygen covered region in Fig. \ref{Figure2}
(f), we also performed finite temperature, ab-initio molecular dynamics calculations.
The Nose thermostat is used and the time steps are taken 2.5 femtoseconds. Atomic velocities are
normalized at every 40 time steps and calculations lasted for 10 picoseconds at
1000 K for supercell containing 128 carbon atom and 10 oxygen adatoms. At the end, this
structure remained stable and neither $O_{2}$ formation, nor dissociation of oxygen atoms
from graphene surface did occurred. High binding energies of adsorbed oxygen atoms and absence
of oxygen dissociation or any irreversible structural transition at 1000 K suggest
that the oxygen covered domains shall remain intact for reasonable times in spite of
the fact that underlying graphene is locally distorted.

\subsection{Coverage of oxygen on both sides of graphene }

\begin{figure}
\includegraphics[width=7cm]{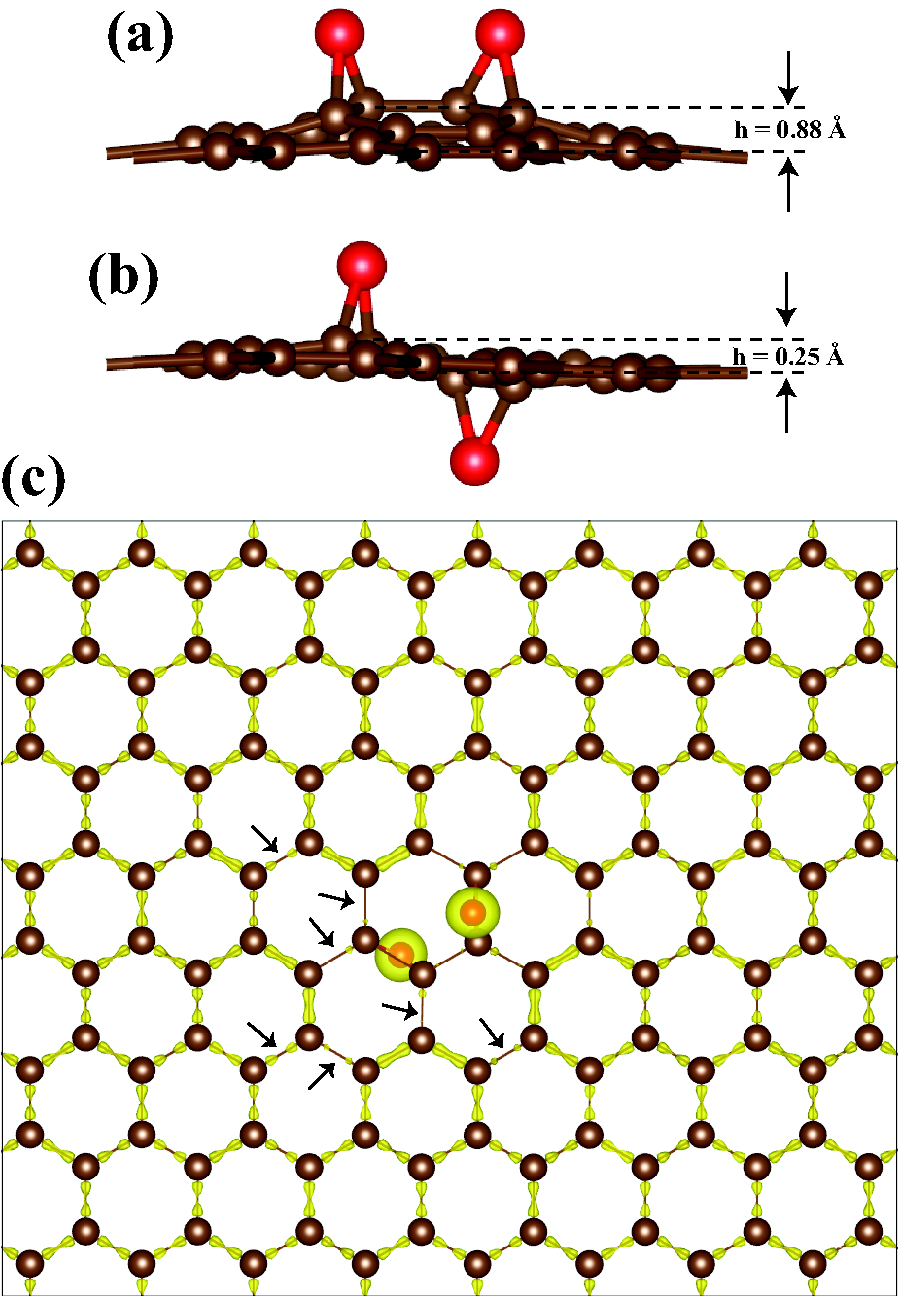}
\caption{(Color Online) Adsorption of two oxygen atoms on one surface of graphene with
buckling of 0.88 \AA. (b) Adsorption of two oxygen atoms at both sides with a buckling of 0.25 \AA. (c)
Isosurfaces of bond charge densities after the adsorption of two oxygen atoms, each one
adsorbed to different sides of graphene. Some of the C-C bonds of graphene, which are deprived of regular charges upon oxidation, are highlighted with arrows.}
\label{Figure_altust}
\end{figure}

The adsorption of oxygen atoms on one side of graphene and hence formation of domain structure
induces structural deformations at the underlying graphene. Fig. \ref{Figure_altust} (a)
shows the side-view of graphene structure with two oxygen atoms adsorbed on the same
surface of graphene. The carbon atoms
below the adsorbed oxygen atoms are distorted and raised towards oxygen atoms. The resulting
buckling is as large as 0.88 \AA{} above the plane of graphene.  The amount of these
distortions are further increased with the addition of more oxygen atoms.
In contrast to this situation, the amount of buckling is reduced to 0.25 \AA,
if the second oxygen atom were adsorbed to the other
surface of graphene, as shown in Fig. \ref{Figure_altust} (b).
Nonetheless, the latter configuration is 110 meV more energetic than the previous
case. This result is good in agreement with Ref[\onlinecite{yan-prb}]. Hence, the two-sided
adsorption shall be preferred instead of the single-sided adsorption.
Despite that the favorable binding site of second oxygen atom at the other side follows our
bond charge density analysis discussed in Sec. IV (B). For example, when
an oxygen atom is adsorbed on one side as in Fig. \ref{Figure2} (a), the most
favorable adsorption side for second oxygen is the same and is $B_{1-2}$ site no 
matter whether the second oxygen adsorbs to the top surface (one-sided adsorption) 
or to the bottom surface (two-sided adsorption). Moreover, the isosurface 
charge density profiles shown in Fig. \ref{Figure_altust} (c)
for the case of second oxygen adsorbed on other side are identical to the profile
in Fig. \ref{Figure2} (b) when two of the oxygens are adsorbed on one side. The ordering
of energetically favorable sites presented in Fig. \ref{Figure2} for higher oxygen
coverage is independent of the adsorption side. Nonetheless, two sided adsorption is
energetically more favorable.

\subsection{ Carbon (C) and Fluorine (F) adsorption on graphene }

\begin{figure}
\includegraphics[width=8cm]{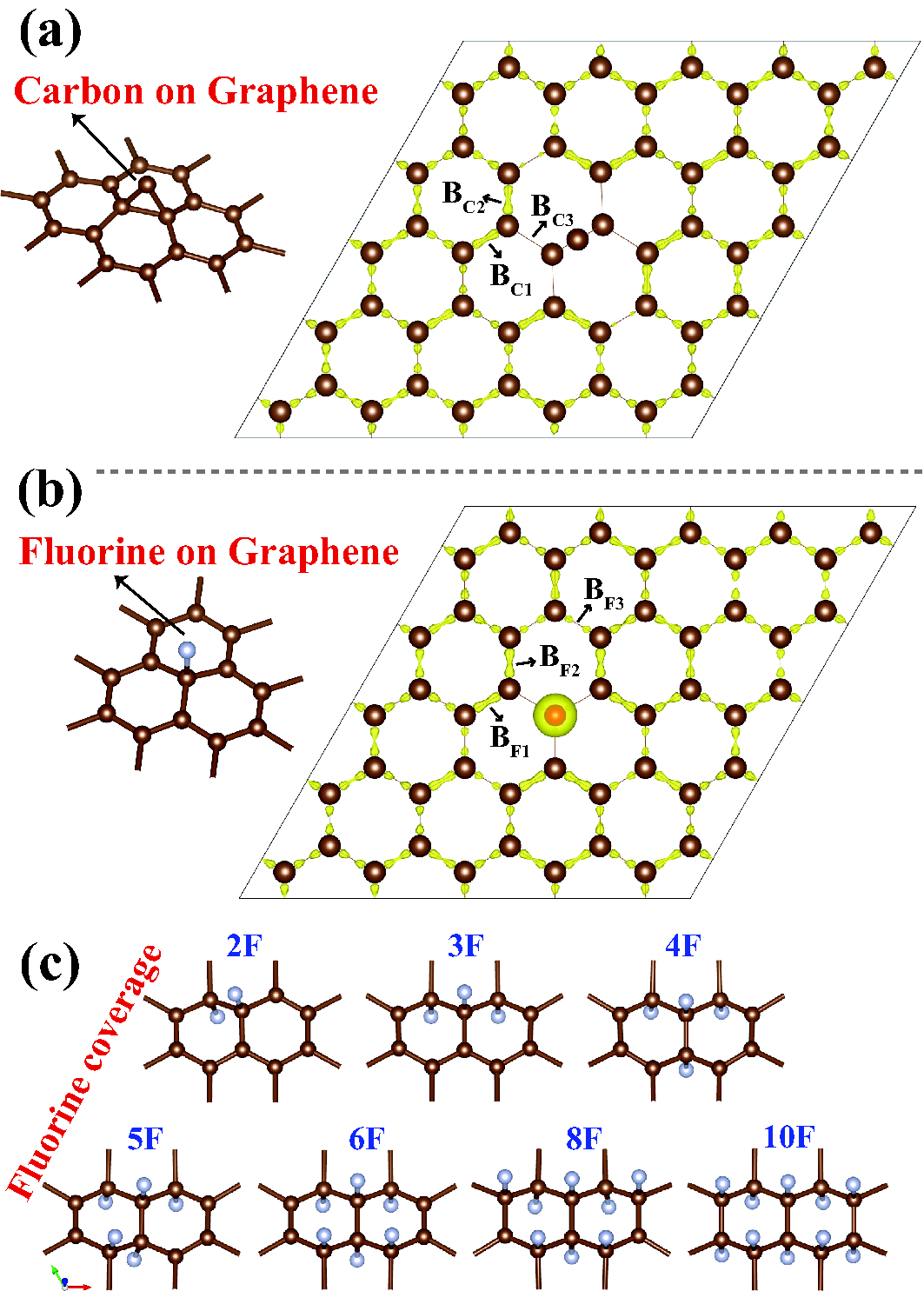}
\caption{(Color Online) (a) The bonding configuration of a single carbon atom on graphene
and the resulting redistribution of bond charges shown by isosurfaces. (b) The bonding
configuration of a single fluorine adatom on graphene with energetically favorable top
site. The resulting charges of C-C bonds at close proximity are shown by isosurfaces.
(c) The growth pattern in the course of the fluorination of graphene.}
\label{Figure_CF}
\end{figure}

We now investigate the adsorption of carbon (C) and
fluorine (F) atoms on graphene in the context of previous charge density
analysis. Similar to oxygen atom, carbon atom is adsorbed at the bridge
site. The atomic structure of carbon atom adsorbed on graphene presented
in Fig. \ref{Figure_CF} (a) is reminiscent of the oxygen atom adsorption
on graphene as presented in Fig. \ref{Figure1} (a). The distance between the adsorbed
carbon atom and nearest-neighbor carbon atoms of graphene is 1.52 \AA, which is
slightly larger than the distance of nearest carbon-oxygen
atoms (1.46 \AA) in GOX. The angle formed between adsorbates and host
graphene atoms which is 62.5 degree is almost equal for carbon and oxygen
adsorption. On the other hand,  the binding energy of carbon atom adsorbed
on (5x5) supercell of graphene is 1.56 eV and significantly smaller than the
binding of oxygen adatom on graphene. The Bader analysis calculates a charge
transfer of 0.04 electrons from the adsorbed carbon atom at the bridge site to
the host graphene atoms and this value is also significantly smaller and
is in the reverse direction as compared charge transfer between carbon and
oxygen atoms. Consequently, the chemical interaction between carbon atoms
is covalent rather than ionic. Nonetheless, owing to formation of new
covalent bonds between C adatom and graphene the isosurface charge density of C-C bonds
presented in Fig. \ref{Figure_CF} (a) mimics the isosurface in the case of
oxygen adsorbed on graphene as presented in Fig. \ref{Figure1} (d). The
nearby bonds of $B_{C1}$ and $B_{C2}$ contain more electronic charge compared
to $B_{C3}$ and other C-C bonds as shown in Fig. \ref{Figure_CF} (a). Moreover,
it was argued that local disturbances on graphene are long ranged.\cite{longrange,bacsi,lehtinen,can}
Interestingly, when a second carbon atom is adsorbed at the close proximity of a second
carbon adatom, the bonds of the first one are broken and subsequently it is attached on top
of the second adsorbed C atom to form C$_2$ molecule. This way $\sim$ 5 eV energy is gained.
The growth of C$_n$ atomic chain continues whenever an adsorbed carbon atom approaches
the existing C$_{n-1}$ chain, whereby the chain is detached from graphene and
attached to the top of adsorbed carbon atom. These results confirm the earlier study
on the perpendicular growth of C$_n$ chains on graphene.\cite{can_chain} While attractive interaction between
adsorbed carbon atoms on graphene give rise to the growth of chains on graphene, the
repulsive interaction between oxygen adatoms hinders the formation of O$_2$ molecules.

The atomic structure and isosurface charge density
profile of fluorine atom adsorbed on graphene is presented in Fig. \ref{Figure_CF} (b).
For the case of fluorine adsorption, the energetically favorable site is the top site as
shown in Fig. \ref{Figure_CF} (b). The binding energy of single F atom adsorbed
on a (4x4) graphene is calculated within LDA and was found to be 2.71 eV.\cite{CF2}
However, present calculations using PBE+vdW correction\cite{grimme} yield a binding energy
of 1.99 eV for adsorption of single F atom on a (5x5) graphene supercell.
Upon fluorine adsorption, the bond charge of nearby atoms is modified as shown in
isosurface profile. The Bader analysis yields a charge transfer of 0.57
electrons from carbon atoms to the adsorbed fluorine atom at the top site and this
value is also significantly close to the value of charge transfer
between carbon and oxygen atoms in the present study. Similar to C and O adsorption, the
nearby bonds of $B_{F1}$ and $B_{F2}$ contain more electronic charge compared
to  $B_{F3}$ and other C-C bonds. The nearest top site between $B_{F1}$ and
$B_{F2}$ bonds and its other two analogues around F atoms contain more
electronic charge and it turns out that these are energetically
most favorable sites for adsorption of additional F atoms.

In Fig. \ref{Figure_CF} (c) we present how F atoms cover graphene. The second F atom
is bonded to the top site formed by $B_{F1}$ and $B_{F2}$ bonds at the other side of graphene, which
is most favorable site compared to to others. The third and fourth F atoms are also bound to other two analogues of this site. The energetics of binding structure are in complete
agreement with the amount of bond charges of nearby top sites. The final arrangement
containing 10 F atoms show a well defined pattern and further fluorination will be
continuation of this pattern.

\section{ Oxygen - Oxygen Interaction }

\begin{figure}
\includegraphics[width=8cm]{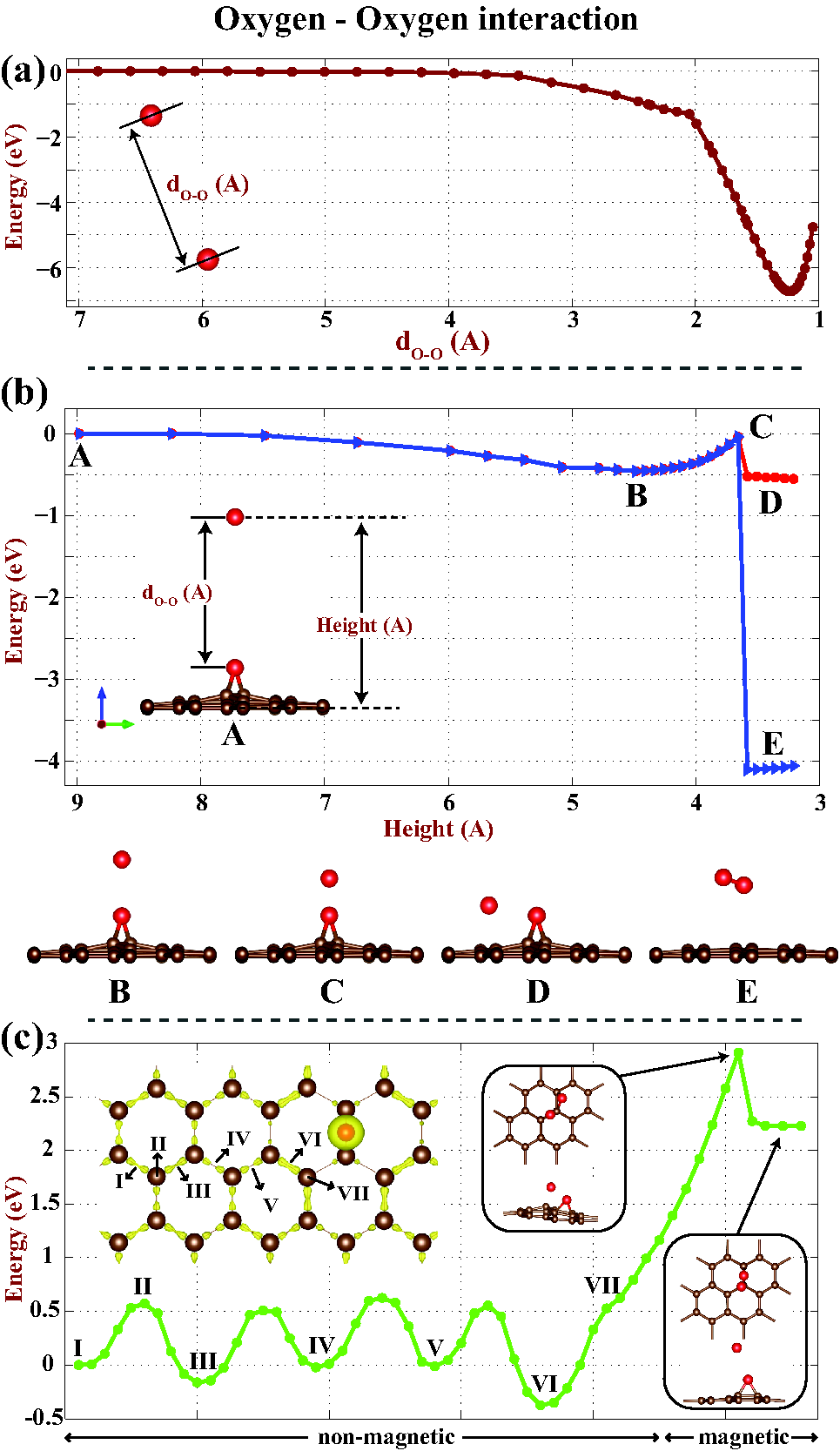}
\caption{(Color Online) (a) The interaction energy between two free oxygen atoms approaching
each other from a distance. The distance between them is $d_{O-O}$. (b) The interaction
energy between a single oxygen atom adsorbed at the bridge site and a free oxygen atom
approaching from the top. Different positions of approaching O atom are shown in the
side views. (c) Variation of the energy between two oxygen adatoms on graphene.
Some of the positions of the approaching oxygen atom on the path of minimum energy
barrier are labeled by numerals ( I-VII ). Top and side views of the
configuration of two oxygen atoms are shown by insets.}
\label{Figure_O-O}
\end{figure}

The interaction between two free oxygen atoms in vacuum is attractive and the formation
of an oxygen molecule is energetically more favorable. We set the total energy to zero when the
distance $d_{O-O}$ between them is 7 \AA. Figure \ref{Figure_O-O} (a)
shows the variation of the energy with the distance, $d_{O-O}$, between two
oxygen atoms. The energy does not vary until $d_{O-O}$ is 3.5 \AA, but it starts to
decrease as $d_{O-O}$ decreases and passes through a minimum for $d_{O-O}=1.21$ \AA~.
This minimum corresponds to the equilibrium bond length of O$_2$ molecule with a binding
energy of 6.67 eV. The process is exothermic and occurs without any energy barrier. However, the
situation is rather different when one of the oxygen atom is adsorbed to the graphene
surface and the other one is free, but approaching from above towards it. In this case, the
position of free oxygen is fixed at preset heights while it is approaching, the rest of
the system consisting of adsorbed oxygen and all graphene atoms are fully relaxed within
conjugate gradient method. We label some of the stages by letters, A-B-C-D-E, while
the two oxygens are approaching each other as shown in Fig. \ref{Figure_O-O} (b). The O-O
coupling is initially negligible at large $d_{O-O}$ at A, but it passes through a minimum by
lowering 0.5 eV at point B corresponding to $d_{O-O} = 2.63$ \AA. Further decrease of
$d_{O-O}$ increases the energy increases until the point C, which is $\sim$ 0.5 eV above
the point B. Beyond C, oxygen atom flips sideways at D. If one prevents oxygen atom from
flipping by fixing its \textit{x-y}-position, but forces it towards the oxygen atom adsorbed
on graphene, the adsorbed one is desorbed and two oxygen atoms form O$_2$ molecule at E.
In this exothermic process, once the barrier is overcame, the energy decreases by $\sim$ 3.5 eV.

For the case of two oxygen adatoms, both adsorbed on graphene and approaching towards each other,
the variation of interaction energy is given in Fig. \ref{Figure_O-O} (c).
Some of the positions of the approaching oxygen atom on the path of minimum energy
barrier are  labeled by numerals. For an oxygen starting from a bridge site at I
and approaching towards the other oxygen, the energy shows an oscillatory behavior.
The maxima, such as II, correspond to the positions where the approaching oxygen is at top site,
while the minima, such as I, III, IV, V, VI correspond to positions at the bridge site.
The charges of bond charge at the bridge site for reasons discussed before result in the changes in
the energies at the bridge sites. For example, the bridge site VI contains more
electronic charge and hence it marks the lowest energy position as one oxygen adatom is approaching
the other oxygen adatom. The energetics of diffusion through the path between V and VI and
energy barrier between them is in good agreement with the calculations by Sun and Fabris.\cite{sun}
Since our objective is to investigate the desorption of adsorbed oxygen from graphene surface,
we did not consider the energetics of diffusion from site VI to the bridge position on the C-C bond,
which is parallel to the C-C bond holding the other oxygen. However, in Ref[\onlinecite{sun}]
the barrier to jump to this site is higher.

Beyond the point VII, the non-magnetic oxygen/graphene system acquires
finite magnetic moments of $\approx$ 0.3 $\mu_B$. Due to the repulsive interaction 
between two oxygen atoms adsorbed on graphene the energy increases by 
$\sim$3.2 eV as shown in the Fig. \ref{Figure_O-O} (c). Eventually, the 
approaching oxygen atom is released from the graphene when the energy 
barrier is overcame. The final structure is shown by inset. These results 
indicate that the binding energy of each oxygen on graphene is quite
strong and the formation of oxygen molecule as a result of two oxygen 
atom approaching each other requires significant energy barrier to overcome.

\section{ Electronic properties varying with oxygen coverage }

\begin{figure}
\includegraphics[width=8cm]{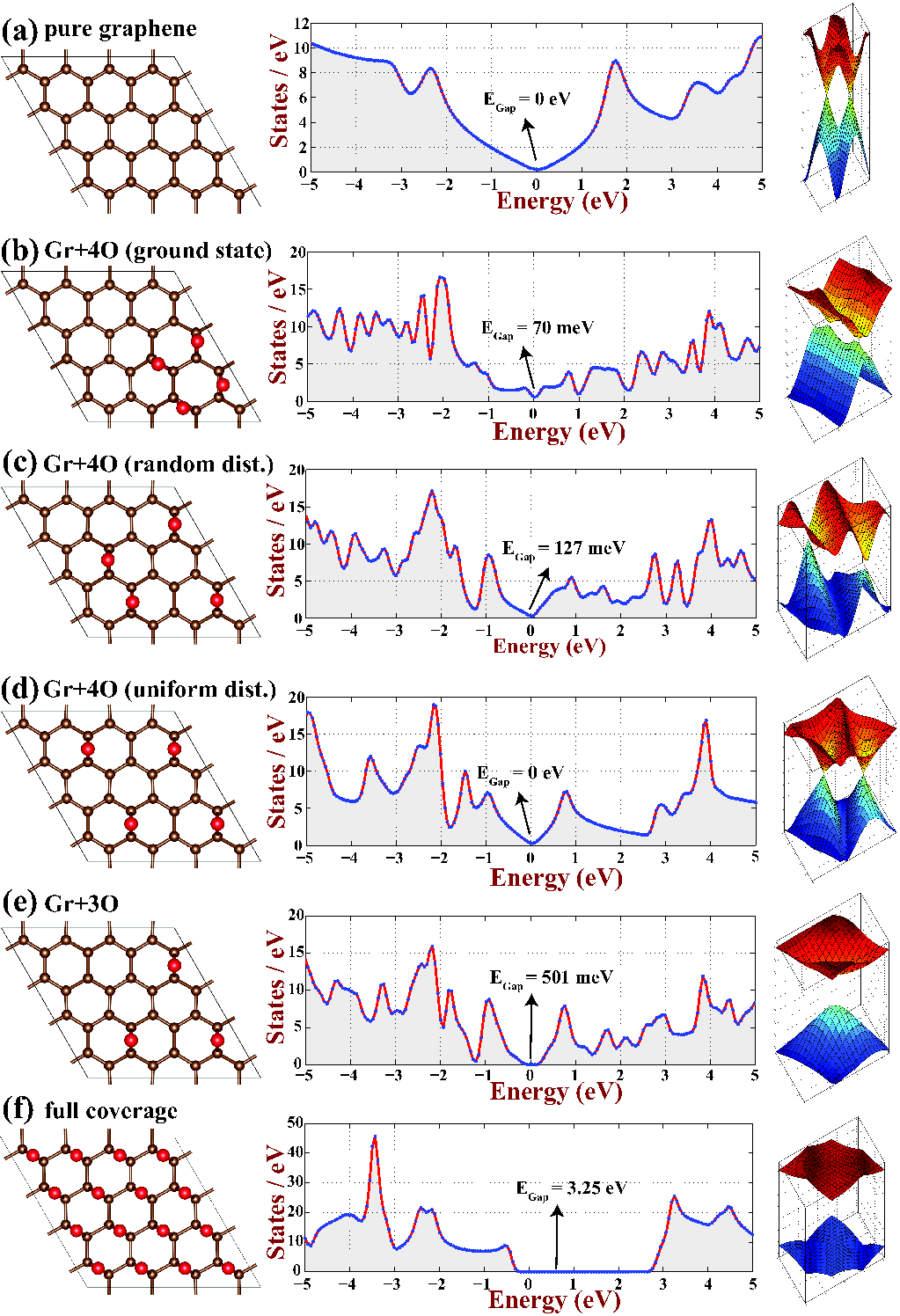}
\caption{(Color Online)  (a) Bare graphene and its typical density of states with zero
state density at the Fermi level E$_F$. The constant energy surfaces of conduction and valence bands are shown on left-hand side. (b) A four-adatom domain corresponding to lowest total energy configuration and has a band gap of 70 meV. (c) Another adsorption configuration of four oxygen adatom resulting in a band gap of 127 meV. (d) A uniform and symmetric configuration of adsorbed oxygen atoms with zero density of states at $E_F$. (e) A sizable band gap is opened when the symmetry of oxygen decoration is broken by the removal of a single oxygen atom. (f) The wide band gap of 3.25 eV is opened for coverage corresponding to $\Theta = N_{O}/N_{C} = 0.5$ at one side (N$_O$ and N$_C$ are the numbers of oxygen and carbon atoms in the (4x4) supercell.}
\label{Figure3}
\end{figure}

The electronic energy structure of GOX strongly depends on oxygen coverage, as well
as on the pattern of coverage. Here we consider the electronic properties corresponding
to different number of oxygen atoms adsorbed at different bridge sites of the (4x4) supercell
repeating periodically. Bare graphene has a semimetallic electronic structure with its
characteristic density of states (DOS) making a dip at the Fermi level and linearly crossing valance
and conduction bands at special K- and K'-points of Brillouin Zone as shown
in Fig. \ref{Figure3} (a). It has a zero band gap and these special symmetry points
are called Dirac points.\cite{dirac} The energetically favorable configuration
of four oxygen atoms adsorbed on a (4x4) hexagonal supercell is presented in
Fig. \ref{Figure3} (b). The resulting DOS profile is different from the bare graphene,
since a narrow energy gap of 70 meV is opened. The Dirac cones disappeared
and the band gap occurs at the points different than K- and K'-points. For a
random and energetically less favorable distribution of oxygen atoms as
in Fig. \ref{Figure3} (c), the energy band gap is further increased to
127 meV. The position of the minimum of conduction band and the maximum of
the valence band has changed in BZ. Surprisingly, the semimetallic band structure of
graphene is recovered when four oxygen atoms are uniformly disturbed on
graphene surface as shown in Fig. \ref{Figure3} (d). Although the difference
of the atomic positions from Fig. \ref{Figure3} (c) is minute, the band gap
is closed and the density of states profile becomes similar to that of bare graphene
making a dip at the Fermi level. The conduction and valance bands cross at
K- and K'-special points similar to bare graphene.

Earlier, it was reported that the superstructures and nanomeshes having
special point-group symmetry, which are generated by decoration of adatoms,
adatom groups or holes repeating periodically in graphene matrix may give
rise to the linearly crossing bands and hence to the recovery of massless
Dirac Fermion behavior.\cite{hasan-nanomesh} Our results
in Fig. \ref{Figure3} (d) is a verification of this situation for oxygen
adsorption on graphene. However, the perfect uniform coverage of oxygen
on graphene is experimentally not achievable and we also present a situation
where the periodicity of uniform coverage is broken by removal of an oxygen
atom as in Fig. \ref{Figure3} (e). In contrast to electronic structure as
in Fig. \ref{Figure3} (d), the Dirac behavior is completely removed and
the resulting structure is semiconductor with relatively large energy band
gap of 501 meV.

For the case of $N_{O}/N_{C}$=0.5 coverage at one side, the resulting
system is a wide band gap material. Fig. \ref{Figure3} (f) shows the atomic
structure and density of states profile. Unlike bare graphene and low oxygen
coverage, the resulting structure has a band gap of 3.25 eV. The two sided
coverage for $N_{O}/N_{C}$=0.5 also yields similar DOS profile with a band gap wider than 3 eV.
These results indicate that the regions of GOX where each carbon atom is
bonded with an oxygen should be an insulator and hence should reflect light.

\section{ Discussions and Conclusions }
Our study dealt with the adsorption of single and multiple
oxygen atoms to graphene surface and explored their desorption.
We showed that O$_2$ molecule can merely be physisorbed to graphene surface.
In contrast, free oxygen atoms are adsorbed at the bridge sites above C-C bonds
by forming strong chemical bonds. Significant amount of charge is
transferred to oxygen adatom from graphene, which disturbs the charge distribution
of the C-C bonds at the proximity of adsorbate. Additional oxygen atoms are adsorbed
to the bridge sites above the C-C bonds of graphene, which has highest charge density.
This behavior promotes the developments of domains of oxygen adatoms. The domain
pattern which, in fact is energetically favorable is also preserved for oxygen atoms
adsorbed to both sides of graphene. The binding  energy of adsorbed oxygen atoms
display an oscillatory change; it starts from 2.43 eV and eventually raises
to 2.80 (3.34) eV at one sided (two sided) full coverage with $\Theta$=0.5.
Accordingly, the formation energy of adsorbed oxygen is negative.

Even if the sequential adsorption of oxygen atoms forms domains with
nonuniform coverage, full coverage can form eventually. High oxygen 
coverage only at one side of graphene causes to severe deformations of 
graphene lattice. While the adsorption configurations which can be precursors of
unzipped of graphene are not favorable for low coverage of large graphene 
surfaces, they may occur at the edges of domains comprising large number 
of oxygen atoms, where underlying graphene lattice is severely distorted.
Nonequilibrium conditions occurring at finite temperatures and size effects 
originating from the small size of underlying graphene may favor the nucleation
of precursors of unzipping. 

Single oxygen migrates on a pathway of minimum energy barrier of 0.6 eV over
the honeycomb structure between bridge and top sites. For the same reason the
interaction between two oxygen adatoms exhibits an oscillatory variation, but
becomes increasingly repulsive as the distance decreases beyond a threshold value.
This repulsive interaction hinders desorption of oxygen through the formation
of O$_2$ molecule despite the negative formation energy of adsorbed oxygen atoms.

The electronic structure of oxidized graphene is strongly dependent on the coverage
of oxygen and its configuration. While the massless Dirac Fermion behavior with
linearly crossing bands at the Fermi level is maintained for specific coverage conserving
certain rotation symmetry, the band gap opens and develops with increasing coverage
of oxygen adatom. As oxidized domains dominate the surface, semimetallic bare graphene is
transformed into a semiconducting material. It appears that the band gap
can be engineered through oxygen coverage. Bright and dark spots observed experimentally
on GOX surfaces are expected to be related with metallic and light reflecting semiconducting
regions, respectively. We believe that metallic regions corresponds to $sp^2$-bonding
regions of graphene. Our results indicate that a specific external effect is required for 
the fast and reversible transition between metallic and semiconducting states of graphene oxide.

\section{ Acknowledgements }
This work is supported by TUBITAK through Grant No:108T234. All the
computational resources have been provided by TUBITAK ULAKBIM, High
Performance and Grid Computing Center (TR-Grid e-Infrastructure).
S. C. acknowledges the partial support of TUBA, Academy of Science of
Turkey.

\end{document}